\documentclass[twocolumn,aps,showpacs,showkeys,preprintnumbers,amsmath,amssymb]{revtex4}

\usepackage{graphicx}
\def\equationautorefname#1#2\null{Eq.#1(#2\null)}

\usepackage{amsmath}
\usepackage{amssymb}
\usepackage{amsthm}
\usepackage{hyperref}
\newcommand{\ii}{\mathrm{i}}
\newcommand\bovermat[2]{%
  \makebox[-6pt][c]{$\smash{\overbrace{\phantom{%
    \begin{matrix}#2\end{matrix}}}^{\text{#1}}}$}#2}

\begin{document}

\title{Correlation energy for the homogeneous electron gas: exact Bethe-Salpeter solution and new approximate evaluation}

\author{Emanuele Maggio}
\author{Georg Kresse} 
\email{georg.kresse@univie.ac.at}

\affiliation{University of Vienna, Faculty of Physics and Center for
Computational Materials Science, Sensengasse 8/12, A-1090 Vienna, Austria }
\date{\today}

\begin{abstract}
\noindent The correlation energy of the homogeneous electron gas is evaluated by solving the Bethe-Salpeter equation (BSE) beyond the Tamm-Dancoff approximation for the electronic polarisation propagator. 
	The BSE is expected to improve upon the random phase approximation, owing to the inclusion of exchange diagrams. 
	For instance, since the BSE reduces in second order to M{\o}ller-Plesset perturbation theory, it is self-interaction free in second order. 
	Results for the correlation energy are compared with Quantum Monte Carlo benchmarks and excellent agreement is observed. 
	For low densities, however, we find imaginary eigenmodes in the polarisation propagator.
	To avoid the occurrence of  imaginary eigenmodes, an approximation to the BSE kernel is proposed, which allows to completely remove this issue in the low electron density region. 
	We refer to this approximation as the random phase approximation with screened exchange (RPAsX). We show that this approximation even slightly improves upon the standard BSE kernel. 
\end{abstract}

\pacs{xxx}

\maketitle

\section{Introduction}
		The study of the homogeneous electron gas (HEG), as a model in condensed matter theory, has a long tradition. 
	In fact, it allows to focus on the properties of the many electron system without complications arising from the discretised lattice symmetry.
	Instead, the neutralising positive charges are uniformly spread out and non-responsive.
	Even if interactions among electrons are not present in the model, its pair correlation function is not constant.
	This is a consequence of the Pauli principle affecting same-spin electrons.
	The exchange energy introduced this way promotes a spin polarisation within the electron gas and, at the same time, increases the electrons' kinetic energy \cite{Giuliani2005}. 
	As the Coulomb repulsion among electrons is turned on, correlations between particles (prevalently of opposite spin) \cite{Endo1999} set in.
	This mechanism produces even stronger deviations from the free-electron case. 
	One of the reasons for a continued interest in this model is that it provides density functional theory with a natural starting point for the unknown exchange and correlation potential.

		An attempt to supersede the mean-field description of the HEG dates back to the seminal work by Hubbard \cite{Hubbard1958}.
	There, a local field factor was introduced to compute the system's response functions.
	Albeit progress has been made to evaluate this quantity, whose asymptotic behaviour is known exactly \cite{Gori-Giorgi2000}, its complete characterisation is still lacking \cite{PhysRevB.62.16474}. 
	On the other hand, the field factor can be related to the irreducible electron-hole scattering amplitude.
	This appears in the integral (and recursive) equation for the polarisation propagator, known as the Bethe-Salpeter equation (BSE). 
	Therefore by solving the BSE one can gain access to exchange and correlation properties.
	The BSE kernel $I$ itself is related to the irreducible self-energy through the relation $I=-\frac{1}{i} \frac{\delta \Sigma}{\delta G}$.
	Hence, the degree of sophistication in the BSE approach can be systematically improved by including more self-energy diagrams at the many-body perturbation theory (MBPT) level. 
	In this study, we implement a computational scheme to solve the BSE when its kernel is derived from the $GW_0$ approximation for the self-energy.
	We then evaluate the correlation energy of the HEG and compare it against Quantum Monte Carlo benchmarks. 

		The $GW_0$ and the Random Phase Approximation (RPA) include topologically equivalent  energy diagrams \cite{Caruso2013b}.
	In both cases, a sub-class of self-energy diagrams, the so-called "bubble" diagrams are summed up to infinite order \cite{Stefanucci2013}. 
	While the RPA approach has witnessed alternate fortune over the years, there has been a strong resurgence of interest in the field recently \cite{Furche2001, Miyake2002,Niquet2003a, Jiang2007, Furche2008, Scuseria2008, Harl2008b, Harl2010a, Toulouse2009a, Toulouse2010a, Jansen2010b}.
	Application of the RPA method to molecules \cite{Furche2001, Olsen2013a, Heßelmann2010a, Macher2014, Teale2010, Bleiziffer2012} 
	and solid state systems \cite{Miyake2002,Harl2010a, Kaltak2014, KresseSOSEX, Schimka2010,Garcia-Gonzalez2007} 
	has highlighted its capabilities but also its limitations. 
	The main advantage of RPA related schemes stems from the seamless inclusion of long-range dispersion (typically not included in standard DFT potentials) \cite{Harl2008b}. 
	On the other hand, systematic underestimation of binding energies \cite{Furche2001, Kresse2009, PhysRevLett.106.153003} 
	and non-physical features for dissociation curves in diatomic systems \cite{Mori-Sanchez2012a, Henderson2010}⁠ are some shortcomings of the RPA.
	To improve on this, the BSE has been exploited 
	to describe hydrogen dissociation \cite{Olsen2014a}, 
	to gain access to optical properties of semiconductors and insulators \cite{Albrecht1998, Rohlfing2000, Yan2012} 
	and to study excitonic effects in extended \cite{Gatti2013a} and molecular systems \cite{DelPuerto2006, Duchemin2013}.
	To the best of our knowledge, the vast majority of previous studies not concerned with exciton characterisation have been carried out on DFT reference states; this is rectified herein.
	
		To assess the correlation energy of the HEG the adiabatic connection (AC) formalism will be employed in this study.
	From its inception, the AC formalism has been applied to obtain the correlation energy starting from reference states evaluated either using Hartree-Fock \cite{Linderberg1967} or Kohn-Sham \cite{Gunnarsson1976, Langreth1977, Harris1974} frameworks.
	In either case, the fluctuation-dissipation theorem is exploited to relate the ground state correlation energy to the system's linear response functions integrated over the AC path. 
	
		The study of the HEG is challenging for many reasons.
	Among others, correlations associated with large momentum transfer interactions \cite{PhysRev.176.589} and between spin-parallel electrons \cite{Endo1999} are overestimated by RPA. 
	The computational scheme employed here addresses these issues by including screening effects and exchange contributions to infinite order in perturbation theory. 

	The manuscript is structured as follows: we first derive an expression for the correlation energy in terms of the polarisation propagator.
	Then in \autoref{sec:BSE} the implementation of the BSE as a non-Hermitian generalised eigenvalue problem is described.
	Properties peculiar to the Bloch representation are discussed at the end of the theory section. 
	Computational details are reported in \autoref{sec:CD}; finally in \autoref{sec:Results} the computational scheme proposed and related approximations are put to fruition to assess the correlation energy of the HEG.

\section{Theory}	

	
\subsection{Exchange and Correlation Energy evaluated along the AC path}
\label{sec:EcDer}

		The adiabatic connection formalism was originally introduced in the DFT framework \cite{Gunnarsson1976, Langreth1977, Harris1974} in order to provide a compact expression for the exchange and correlation energy. 
	In DFT this quantity is given by the sum of two contributions: the electron-electron interaction and the difference in the kinetic energy between the `physical' system and the Kohn-Sham system. 
	These terms are related to the two body density matrix and to the one body density matrix respectively.
	Similarly, we can also decompose the exchange-correlation energy into one- and two-body contributions.
	The two-body part of the correlation energy can be estimated solving the BSE, as described in \autoref{sec:BSE}. 
	The impact of one-body contributions has been the subject of recent studies \cite{Klimes2015, PhysRevLett.106.153003}.
	In this section we compactly derive an expression for the correlation energy and discuss the main assumptions required, the details are reported in Appendix \ref{appx:G2}.

	The system's Hamiltonian in second quantised form along the AC path reads:
        \begin{widetext}
	\begin{equation}
		\label{eq:H2nd}
		\hat{\mathcal{H}}_\lambda= \int d\mathbf{x}d\mathbf{x}' \hat{\psi}^\dagger(\mathbf{x}) \langle \mathbf{x} | \hat{h}| \mathbf{x'} \rangle \hat{\psi}(\mathbf{x}') 
		 + \frac{\lambda}{2} \int  d\mathbf{x}d\mathbf{x}' v(\mathbf{x}, \mathbf{x}')\hat{\psi}^\dagger(\mathbf{x})\hat{\psi}^\dagger(\mathbf{x}')\hat{\psi}(\mathbf{x}')\hat{\psi}(\mathbf{x})+ 
		\int d\mathbf{x}d\mathbf{x}' \hat{\psi}^\dagger(\mathbf{x}) \langle \mathbf{x} | \hat{V}^{\rm{xc}}_\lambda| \mathbf{x'} \rangle \hat{\psi}(\mathbf{x}').
	\end{equation}
        \end{widetext}

	The field operators $\hat{\psi}(\mathbf{x})$ and their adjoint are given in the ordinary Schr\"odinger picture; 
	the Coulomb potential is defined in the spin-space basis as 
	$v(\mathbf{x,x'})= \frac{\delta_{\sigma,\sigma'}}{|\mathbf{r-r'}|} $ 
	with the usual space-time notation $(\mathbf{x},t)=(\mathbf{r}, \sigma, t)$ for the space $(\mathbf{r})$, spin $(\sigma)$ and time $(t)$ variables. The one-particle operator $\hat{V}^{\rm{xc}}_\lambda$ can be any approximate exchange correlation potential, even one that is not diagonal in real space, for instance, an Hermitian energy independent approximation for the self-energy $\Sigma_{\rm {xc}}(\omega)$.

	Along the AC path the many body interactions are progressively switched on, inducing changes in the correlations between electrons previously captured only by the exchange-correlation potential $ \hat{V}^{\rm{xc}}_0$.
	The latter also varies with varying coupling constant and different switching can be realised for $ \hat{V}^{\rm{xc}}_\lambda$. 
	These are either designed to keep the electron density fixed along the AC path \cite{Langreth1977}, or assume a linear dependence on $\lambda $ \cite{Harris1984,VanAggelen2014a}. 
	Here we use the linear switching $ \hat{V}^{\rm{xc}}_\lambda = (1-\lambda) \hat{V}^{\rm{xc}}_0$.

		The correlation energy can be defined in full generality as the difference between the expectation values of the interacting Hamiltonian acting on its ground state and on the ground state of the non-interacting system:
	\begin{equation}
		E_c= \langle \Psi_1 | \hat{\mathcal{H}}_1 | \Psi_1 \rangle - \langle \Psi_0 | \hat{\mathcal{H}}_1 | \Psi_0 \rangle.
	\end{equation}
	If the symmetry of the ground state does not change along the AC path, then $|\Psi_0 \rangle$ evolves into $|\Psi_1 \rangle$ for non-degenerate ground states thanks to the Gell-Mann and Low theorem. 
	It is easy to see that with $E_\lambda \equiv \langle \Psi_\lambda | \mathcal{H}_\lambda| \Psi_\lambda\rangle $, 
	$ \langle \Psi_0 | \mathcal{H}_1 | \Psi_0 \rangle = E_0 + \langle \Psi_0 | \frac{1}{2}v - \hat{V}^{\rm{xc}}_0 | \Psi_0 \rangle$.
	Then the correlation energy can be expressed as (see also \cite{Klimes2015, Toulouse2010a}):
	\begin{align}
		E_c & =  E_1 - E_0 -  \bigg \langle \Psi_0 \left| \frac{\partial{\mathcal{H}_\lambda}}{\partial{\lambda}} \right | \Psi_0 \bigg \rangle \nonumber \\
		& = \int_0 ^1 d\lambda \frac{dE_\lambda}{d\lambda} -\left. \frac{dE_\lambda}{d\lambda} \right |_{\lambda=0},
		\label{eq:Ec}
	\end{align}
	where we have assumed that the Hellmann-Feynman theorem holds in order to set 
	$\frac{\partial}{\partial{\lambda}} \langle \Psi_\lambda \left| \mathcal{H}_\lambda \right | \Psi_\lambda \rangle = 
	\langle \Psi_\lambda \left| \frac{\partial{\mathcal{H}_\lambda}}{\partial{\lambda}} \right | \Psi_\lambda \rangle.  $ 
	The previous equation reproduces the standard expression found in the literature \cite{Toulouse2010a}. 

		The calculation of the correlation energy thus requires to evaluate the term $\frac{dE_\lambda}{d\lambda} \equiv \dot{E}_\lambda$  along the AC path, where the dependence on the coupling constant is retained by the ground state wave function.
	The derivative of the ground state energy can be written as:
	\begin{align}
		\frac{dE_\lambda}{d\lambda}=&\frac{1}{2} \int \int d\mathbf{x}d\mathbf{x}' v( \mathbf{x}, \mathbf{x}') 
		\langle \psi^\dagger (\mathbf{x}) \psi^\dagger (\mathbf{x}') \psi (\mathbf{x}') \psi (\mathbf{x}) \rangle_\lambda \nonumber \\
		& \label{eq:dEdl}
		- \int \int d\mathbf{x}d\mathbf{x}' \langle \mathbf{x}' | \hat{V}^{\rm{xc}}_0 | \mathbf{x} \rangle \langle \psi^\dagger (\mathbf{x}') \psi(\mathbf{x}) \rangle_\lambda.
	\end{align}
	The last term in the previous equation represents contributions related to the change of the one-particle Green's function to the correlation energy \cite{Klimes2015, PhysRevLett.106.153003}. 
	In the present study, we disregard this term and concentrate on the first term on the r.h.s. of \autoref{eq:dEdl}.
	In fact, in second order, the last term is exactly zero for the HEG
(see \textit{e.g.} chapter 22 in Ref.\onlinecite{Gross1991}), suggesting that its contribution should be small. 

	It is shown in appendix \ref{appx:G2} that the expression above can be recast in terms of the four point polarisation propagator $P_\lambda$ \cite{Toyoda1975}, fulfilling the BSE as detailed in the next section. The correlation energy then reads:
	\begin{equation}
 	\begin{split}
		E_c & = -\frac{1}{2} \int_0^1 d\lambda \int \frac{d\omega}{2\pi} \int \prod_{i=1}^4 d\mathbf{x}_i v(\mathbf{x}_1,\mathbf{x}_4)  \\
	 &	\delta(\mathbf{x}_4-\mathbf{x}_2)\delta(\mathbf{x}_3-\mathbf{x}_1)  \left [ P_\lambda ( \mathbf{x}_1,\mathbf{x}_2, \mathbf{x}_3, \mathbf{x}_4; \omega) - \right. \\
		& \left. -P_0 ( \mathbf{x}_1, \mathbf{x}_2, \mathbf{x}_3, \mathbf{x}_4; \omega) \right ].\\
		\label{eq:EcP}
	\end{split}
	\end{equation}
	With the aid of the polarisation propagator's spectral representation obtained in appendix \ref{appx:deriv}, it is possible to carry out the frequency integration analytically, if the inter-particle interaction is not frequency dependent \cite{Fetter1971}. 

\subsection{Bethe-Salpeter equation along the AC path}
\label{sec:BSE}

	The Bethe-Salpeter equation 	$P= {P}_0+{P}_0 I P $ relates
	the fully interacting polarisation propagator $P$ to the independent particle-hole propagator $P_0$, 
	invoking an interaction kernel $I$. The kernel describes all possible interactions between particles and holes \cite{Strinati1988}.
	Solving this equation is generally a formidable task, which warrants the use of approximations. 
	A commonly used one is to make the interaction kernel instantaneous and to include only Hartree and screened exchange interactions in the kernel $I(1,2,3,4)= V(1,2,3,4)-W_0(2,1,3,4)$ \cite{Albrecht1998, Rohlfing2000, Hanke1979, Hanke1980}. 
	Using the conventional space-time notation (where the superscript '+' denotes positive infinitesimal time shifts), one can write:
	\begin{align}
		\label{eq:V}
		V(1,2,3,4)= v(1,4^+)\delta(4,2^+)\delta(3,1^+) \\ 
		\label{eq:W}
		W_0(1,2,3,4)= w(1,4^+)\delta(4,2^+)\delta(3,1^+).
	 \end{align}
	Here $V$ and $W_0$ are the four-point generalisations of the bare Coulomb ($v$) and screened ($w$) interactions, and the shorthand $(i)=({\bf r}_i, \sigma_i, t_i)$ refers to space, spin and time degrees of freedom.
	The adiabatic switching of many-body interactions (presented in the previous section) requires the irreducible kernel to be linearly scaled by the coupling constant $\lambda$ and, correspondingly, the polarisation propagator $P_\lambda$ to be evaluated for each point along the AC path.
	The resulting BSE will then be parametrically dependent on two quantities: the coupling constant $\lambda$ and the frequency variable $\omega$, provided that the kernel is static. 
	This approximation for the kernel implies that $ w(\mathbf{x}_1,\mathbf{x}_2)$ in \autoref{eq:W} is evaluated as 
	$w(\mathbf{x}_1,\mathbf{x}_2)=\{v^{\frac{1}{2}} \cdot \epsilon^{-1}(\omega=0) \cdot v^{\frac{1}{2}}\}(\mathbf{x}_1,\mathbf{x}_2)$, 
	where internal space variables are integrated over and the dielectric function in this study is computed at the RPA level.

		It is convenient to introduce the `particle-hole' basis: $\Phi_M^R(\mathbf{x,x'})=\psi_a(\mathbf{x})\psi^*_i(\mathbf{x}')$ and $\Phi_M^A(\mathbf{x,x'})=\psi_i(\mathbf{x})\psi^*_a(\mathbf{x}')$, 
	for the resonant $(R)$ and antiresonant $(A)$ component labelled by the ``superindex'' $M\equiv(i,a)$.  
		The BSE can then be recast in this basis to yield:  
	\begin{align}
		\mathbf{P}_\lambda(\omega) = \mathbf{P}_0(\omega)+\mathbf{P}_0(\omega)* \mathbf{I}_\lambda (\omega=0)*\mathbf{P}_\lambda(\omega). 
		\label{eq:Pi}
	\end{align}
	The symbol `$*$' indicates the usual matrix product in the chosen orbital basis.
	The independent particle-hole propagator is diagonal in this basis and can be expressed as:
	\begin{align}
		P_0(\mathbf{x}_1,&\mathbf{x}_2,\mathbf{x}_1',\mathbf{x}_2'; \omega)= \operatorname{tr} \left \{ \mathbf{P}_0 (\omega) \right \}= \nonumber \\ 
		& \sum_{i,a} f_i \bar {f}_a \left(\dfrac{\psi_a(\mathbf{x}_1)\psi^*_i(\mathbf{x}_1')\psi_i(\mathbf{x}_2)\psi^*_a(\mathbf{x}_2')}{\omega-\Delta E_{ia}} + \right.  \nonumber \\
		&\left. +\dfrac{\psi_i(\mathbf{x}_1)\psi^*_a(\mathbf{x}_1')\psi_a(\mathbf{x}_2)\psi^*_i(\mathbf{x}_2')}{-\omega-\Delta E_{ia}}	\right),
		\label{eq:L0SR}
	\end{align}
	where $f_n$ is the (fractional) occupancy for the orbital $n$ and $\bar{f}_n=1-f_n$, the indices $i$ and $a$ in the summations above go over the occupied and unoccupied single particle states respectively.  
	$\Delta E_{ia}= (\epsilon_a-\epsilon_i)$ is the independent particle energy difference.  

\begin {figure*} [t!]
		\centering
		\includegraphics [width=14.5cm] {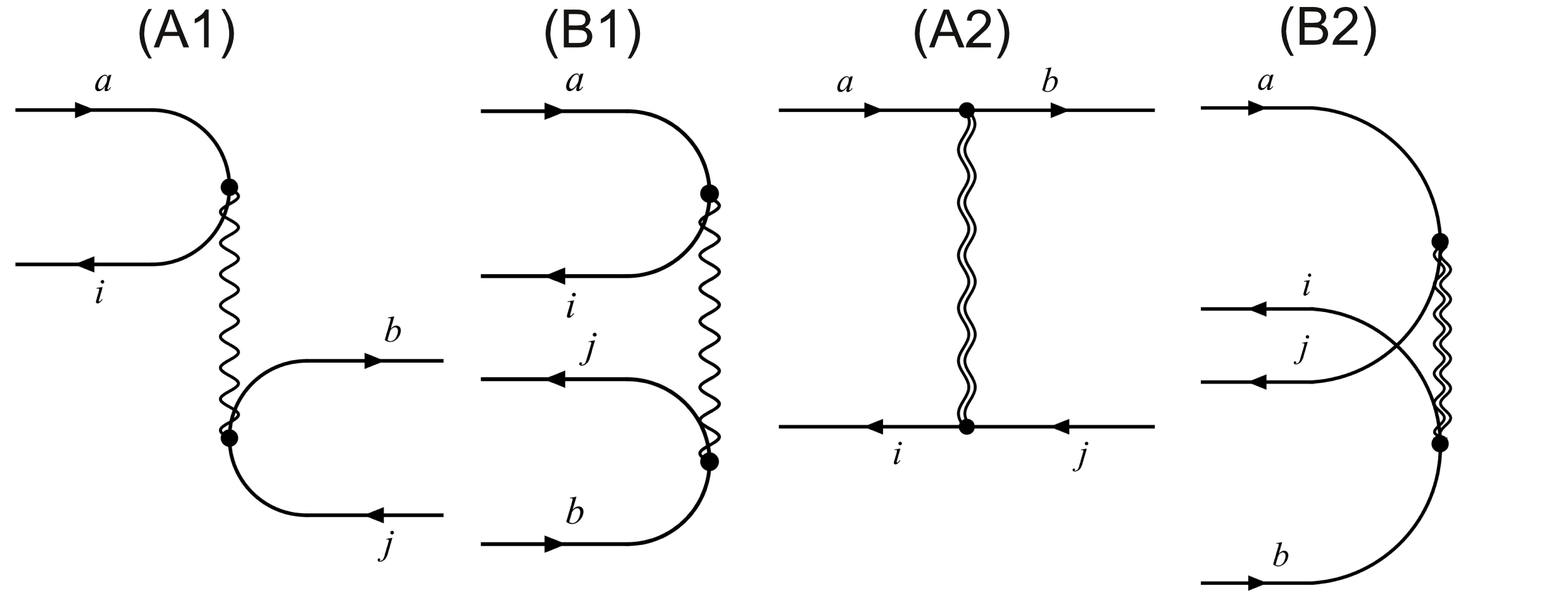}	
		\caption{Goldstone diagrams for the particle-hole interaction included in the BSE kernel: particle states are shown as right arrows $\rightarrow$ (labelled by unoccupied ground state orbitals, $a,b$) and holes by left arrows $\leftarrow$ (labelled by occupied orbitals $i,j$).
	The bare and screened Coulomb interaction is represented by a simple wavy line and a double wavy line respectively.
 	Diagrams (A1) and (B1) represent the resonant and antiresonant particle-hole annihilation and simultaneous creation, whereas diagrams (A2) and (B2) represent the resonant and antiresonant particle-hole scattering process.}
		\label{Fig:BSEkernel}
	\end{figure*}

		Since the BSE kernel is static, it is possible to invert the matrix equation \autoref{eq:Pi} for each frequency point. 
	In appendix \ref{appx:deriv} it is shown that this is, however, not necessary since the spectral decomposition for the polarisation propagator $\mathbf{P}_\lambda(\omega)$ can be constructed solving the non-Hermitian generalised eigenvalue problem (EVP): 
		\begin{align}
		\label{eq:RPA}
		\begin{pmatrix}
			 \mathbf{A}   & \mathbf{B}\\
			 \mathbf{B}^* & \mathbf{A}^*
		\end{pmatrix}
		\begin{pmatrix}
		 	 \mathbf{X}\\
			 \mathbf{Y} 
		\end{pmatrix}= 
		\begin{pmatrix}
			 \mathbf{\Omega} & \mathbf{0}\\
			 \mathbf{0} & -\mathbf{\Omega}
		\end{pmatrix}
		\begin{pmatrix}
			 \mathbf{X} \\
			 \mathbf{Y}
		\end{pmatrix}, 
		\end{align}
	which does not depend on $ \omega$.
	The matrix elements for $\mathbf{A}$ and $\mathbf{B}$ are given by \cite{Fuchs2008}:
	\begin{align}
		A_{iajb}=& (\Delta E_{ia}) + \lambda \langle aj|V|ib \rangle - \lambda \langle aj|W|bi \rangle, \label{eq:RR}\\
		B_{iajb}=& \lambda \langle ab|V|ij \rangle - \lambda \langle ab|W|ji \rangle. \label{eq:RA}
	\end{align}
 	Their time-ordered diagrammatic representation is shown in \autoref{Fig:BSEkernel}.

		The dimension of this non-Hermitian eigenvalue problem is given by the sum of particle-hole and hole-particle transitions. 
	It is possible, however, to reduce the matrix size to a single particle-hole subspace by exploiting the time-reversal symmetry of the system. 
	This statement will be proved in the next section where we introduce a suitable basis for the matrix representation as in Ref. \cite{Sander2015}. 

\subsection {Bloch representation}
\label{sec:plwaves}
		For extended systems it is convenient to switch to a reciprocal space representation of the electronic problem, in order to capture more readily its translational invariance.
	A fully homogeneous system has the highest translational symmetry, which translates into a spherical Fermi surface.
	In this study, we choose to discretise the infinite, translational invariant system into a unit cell of given symmetry and subject to periodic boundary conditions.
	This approach is akin to the QMC benchmarks against which we will compare the outcome of our calculations.
	As a consequence, this discretisation changes the Fermi surface, which is now replaced by a convex polyhedron: this tends to the exact Fermi surface as the number of unit cell replicas (\textit{i.e.} $\mathbf{k}$-point sampling the reciprocal space) in our calculations is increased.
	Each single particle quantum number $a, b, i, j$ can be mapped into a set specifying the particles' band (valence band for the hole states and conduction band for particle states) as well as their Bloch wave-vectors, $ \mathbf{k}, \mathbf{k'} $. 
	We have: 
	\begin{align}
		\label{eq:map}
		\begin{cases}
		a \to (a,\mathbf{k'})  ;  & i \to (i,\mathbf{k'}\pm\mathbf{q})\\
		b \to (b,\mathbf{k})   ;  & j \to (j,\mathbf{k}\pm\mathbf{q})
		\end{cases}
	\end{align}  
	where $\mathbf{q}$ is the momentum of the photon impinging the system and the $\pm$ sign applies to resonant and antiresonant electron-hole pairs respectively.
	We denote the Fourier components of the Coulomb potential with $ v_{\mathbf{q}}(\mathbf{r},\mathbf{r'})$,
	then the two electron integrals in $\mathbf{A}$ are given by integrating over spin $\sigma, \varsigma$ and space variables:
	\begin{align}
		\label{eq:AHartr} \langle aj | V |ib \rangle= \sum_{\substack{\sigma,\sigma' \\ \varsigma,\varsigma'}} \int & d \mathbf{r}d \mathbf{r'}   
		\psi^*_{a,\mathbf{k'}}(\mathbf{r}, \sigma)  \psi^*_{j,\mathbf{k+q}}(\mathbf{r'}, \varsigma') v_{\mathbf{q}}(\mathbf{r},\mathbf{r'}) \times \nonumber \\ & 
		\psi_{i,\mathbf{k'+q}}(\mathbf{r}, \varsigma)  \psi_{b,\mathbf{k}}(\mathbf{r'}, \sigma') \delta_{\sigma,\varsigma} \delta_{\sigma',\varsigma'}\\
		\label{eq:ADirect}\langle aj | W_0 |bi \rangle= \sum_{\substack{\sigma,\sigma' \\ \varsigma,\varsigma'}}\int & d \mathbf{r}d \mathbf{r'}   
		\psi^*_{a,\mathbf{k'}}(\mathbf{r}, \sigma)  \psi^*_{j,\mathbf{k+q}}(\mathbf{r'}, \varsigma') w_{\mathbf{q}}(\mathbf{r},\mathbf{r'}) \times \nonumber \\
		& \psi_{b,\mathbf{k}}(\mathbf{r}, \sigma') \psi_{i,\mathbf{k'+q}}(\mathbf{r'}, \varsigma)  \delta_{\sigma,\sigma'} \delta_{\varsigma,\varsigma'} 
	\end{align}
	and similarly for the $\mathbf{B}$ matrix elements:
	\begin{align}
		\label{eq:BHartr}\langle ab | V |ij \rangle= \sum_{\substack{\sigma,\sigma' \\ \varsigma,\varsigma'}}\int & d \mathbf{r}d \mathbf{r'}   
		\psi^*_{a,\mathbf{k'}}(\mathbf{r}, \sigma)  \psi^*_{b,\mathbf{k}}(\mathbf{r'}, \sigma') v_{\mathbf{q}}(\mathbf{r},\mathbf{r'}) \times \nonumber \\ 
		& \psi_{i,\mathbf{k'+q}}(\mathbf{r}, \varsigma)  \psi_{j,\mathbf{k-q}}(\mathbf{r'}, \varsigma') \delta_{\sigma,\varsigma} \delta_{\sigma',\varsigma'}\\
		\label{eq:BExch}\langle ab | W_0 |ji \rangle= \sum_{\substack{\sigma,\sigma' \\ \varsigma,\varsigma'}}\int & d \mathbf{r}d \mathbf{r'}   
		\psi^*_{a,\mathbf{k'}}(\mathbf{r}, \sigma)  \psi^*_{b,\mathbf{k}}(\mathbf{r'}, \sigma') w_{\mathbf{q}}(\mathbf{r},\mathbf{r'}) \times \nonumber \\
		& \psi_{j,\mathbf{k-q}}(\mathbf{r}, \varsigma') \psi_{i,\mathbf{k'+q}}(\mathbf{r'}, \varsigma)\delta_{\sigma,\varsigma'} \delta_{\sigma',\varsigma}.
	\end{align}
	For the Hartree components in \autoref{eq:AHartr} and \autoref{eq:BHartr} the interaction potential comprises of a single mode $\mathbf{q}$ equal to the difference $\mathbf{k}-\mathbf{k'}$. 
	Whereas, the screened Coulomb interaction in \autoref{eq:ADirect} and \autoref{eq:BExch} can be decomposed in its Fourier modes:
	\begin{align}
		w_{\mathbf{q}}(\mathbf{r,r'})=4\pi e^2\sum_{\mathbf{G,G'}}\frac{e^{-\ii(\mathbf{q+G}) \cdot \mathbf{r'}} \, 
		\epsilon^{-1}_{\mathbf{q+G}, \mathbf{ q+G'}} \, e^{\ii(\mathbf{q+G'}) \cdot \mathbf{r}}}{|\mathbf{q+G}||\mathbf{q+G'}|}.
	\end{align}
		
		The spin structure of the matrix elements in Eqs. (\ref{eq:AHartr}), (\ref{eq:BHartr}) and (\ref{eq:BExch}) spans the singlet subspace, where electron and hole states of each pair have the same spin. 
	The exchange part in the $\mathbf{A}$ matrix (\autoref{eq:ADirect}) is defined in principle also on the triplet subspace where the incoming electron and hole have opposite spin orientations \cite{Rohlfing2000}.
	If the Hamiltonian is time-reversal invariant (\textit{i.e.} spin-orbit coupling is not present) then the singlet and triplet solutions factorise \cite{Rodl2008} and only the former is required to evaluate optical properties and the correlation energy.
 
		We can now move on to prove that, thanks to the time reversal symmetry of the wave function, the original EVP can be downfolded into an Hermitian EVP. 
	The time reversal property of the Schr\"{o}dinger equation implies that, for a single particle state, the Bloch functions observe $\psi_{n,\mathbf{-k}}(\mathbf{r})=\psi^*_{n,\mathbf{k}}(\mathbf{r})$ and the one-electron eigenvalue are also the same. 
	Since the basis functions spanning the space of resonant and antiresonant components are linearly independent, one can selectively invert the single particles' wave-vectors in the antiresonant component as follows:

	\begin{equation}
	\label{eq:mapping}
		\begin{aligned}
			\begin{matrix}
			\\
			\scriptstyle{\mathbf{k} \rightarrow -\bar{\mathbf{k}}}\big \{\\
			\end{matrix}
		\end{aligned}
		\begin{pmatrix}
			 \mathbf{A}   & \bovermat{$\mathbf{k} \rightarrow -\bar{\mathbf{k}}$}{\mathbf{B}}\\
			 \mathbf{B}^* & \mathbf{A}^*
		\end{pmatrix}
		\rightarrow
		\begin{pmatrix}
			 \mathbf{A}   & \mathbf{B}'\\
			 \mathbf{B}' & \mathbf{A}
		\end{pmatrix}.
	\end{equation}
	It is well-known that the $\mathbf{A}$ matrix is Hermitian in this representation \cite{Ring1980}, hence the transformation above will simply map $\mathbf{A}^*$ into $\mathbf{A}$. 
	For the exchange part in \autoref{eq:BExch}, the transformed matrix elements in $\mathbf{B}'$ are given by 
	(with the replacements $\psi_{j,\mathbf{-\bar{k}-q}} \to \psi^*_{j,\mathbf{\bar{k}+q}},  \psi^*_{b,\mathbf{-\bar{k}}} \to \psi_{b, \mathbf{\bar{k}}}$): 
	\begin{align} 
		\label{eq:BMod}
		\int & d \mathbf{r}d \mathbf{r'}  \psi^*_{a,\mathbf{k'}}(\mathbf{r})  \psi^*_{j,\mathbf{k+q}}(\mathbf{r}) w_{\mathbf{q}}(\mathbf{r},\mathbf{r'}) 		 \psi_{b,\mathbf{k}}(\mathbf{r'}) \psi_{i,\mathbf{k'+q}}(\mathbf{r'}).
	\end{align}
	Here, we have dropped the bar over the $\mathbf k$ index and suppressed the spin indices for simplicity. 
	The Hartree component turns out to be identical to the expression in \autoref{eq:AHartr}, hence requiring no extra computational step. 

	Furthermore, the transformation applied to $\mathbf{B}^*$ gives exactly the same matrix elements in \autoref{eq:BMod} as we now show. 
	Starting with the matrix elements in \autoref{eq:BExch} and applying the transformation $\mathbf{k}'\to -\mathbf{\bar{k}}'$ to the antiresonant electron-hole pair $(a, i)$, one obtains: 
	\begin{align}
		\label{eq:BtInv}
		\langle \bar{a} b | W_0 |j \bar{i} \rangle^* = 
		\int d \mathbf{r} d \mathbf{r'} & \psi^*_{j,\mathbf{k+q}}(\mathbf{r})  \psi^*_{i,\mathbf{-\bar{k}'-q}}(\mathbf{r'}) \nonumber \\
		& \times w^*_{\mathbf{q}}(\mathbf{r,r'}) \psi_{a,-\mathbf{\bar{k}'}}(\mathbf{r}) \psi_{b,\mathbf{k}}(\mathbf{r'}). 
	\end{align}
	In the static approximation to the screening potential $w_{\mathbf{q}}(\mathbf{r,r'})$, the dielectric function at the RPA level is an even function of $|\mathbf{q}|$ and has zero imaginary part, hence it is possible to replace $w^*_{\mathbf{q}}(\mathbf{r,r'})$ with $w_{-\mathbf{q}}(\mathbf{r,r'})$ in the equation above.
	Finally, by swapping the integration variables, it is easy to see that the exchange component in $\mathbf{B}^{*'}$ reduces to \autoref{eq:BMod}. 
	The reasoning above can be straightforwardly repeated for the Hartree component in $\mathbf{B}^*$ to give $ \langle a j |V|i b \rangle$. 
	This result completes the proof of \autoref{eq:mapping}. 
	In the following we will drop the prime when referring to the $\mathbf{B}'$ matrix.   

		We can now proceed to reduce the dimensionality of the EVP \eqref{eq:RPA}.
	Adding and subtracting the individual equations contained in \autoref{eq:RPA} and then solving one of the two equations, let's say with respect to $(\mathbf{X-Y})$, one obtains the squared problem \cite{Jørgensen1981}:
		\begin{align}
			\nonumber (\mathbf{A}+\mathbf{B})(\mathbf{A}-\mathbf{B})(\mathbf{X}+\mathbf{Y})= \mathbf{\Omega}^2(\mathbf{X}+\mathbf{Y}).
		\end{align}  
	The expression above can be converted into a conventional EVP if the matrices are positive definite. 
	Then the matrix $(\mathbf{A}-\mathbf{B})^{\frac{1}{2}}$ is unique and the eigenvalue problem finally reads:
		\begin{align}
			\label{eq:RPAsq}
			(\mathbf{A}-\mathbf{B})^{\frac{1}{2}}(\mathbf{A}+\mathbf{B})&(\mathbf{A}-\mathbf{B})^{\frac{1}{2}}(\mathbf{A}-\mathbf{B})^{\frac{1}{2}}(\mathbf{X}+\mathbf{Y})= \nonumber \\
		&= \mathbf{\Omega}^2(\mathbf{A}-\mathbf{B})^{\frac{1}{2}}(\mathbf{X}+\mathbf{Y}).
		\end{align}

		The difference $\mathbf{P}_\lambda(\omega)-\mathbf{P}_0(\omega)$ in  \autoref{eq:EcP} can be integrated over $\omega$ thanks to the spectral decomposition in \autoref{eq:L0SR} and in \autoref{eq:LSR}.
	The correlation energy can then be expressed using the correlation part of the two-particle density matrix $ \mathcal{P}_\lambda$:
	\begin{align}
		E_c=\frac{1}{2} \int_0^1 d\lambda \operatorname{tr} \{ \mathbf{V} \mathcal{P}_\lambda \}
		\label{eq:Ecfinal}
	\end{align}
	with $\mathcal{P}_\lambda=(\mathbf{X}+\mathbf{Y})_\lambda(\mathbf{X}+\mathbf{Y})_\lambda^* - \mathbf{1}$ and $\mathbf{V}$ the Coulomb matrix in \autoref{eq:AHartr} \cite{Furche2001,Angyan2011a}. 	


	\section{Computational details}
	\label{sec:CD}
	
		The HEG is in principle specified by the average electron density $n=\frac{N_e}{V}$ (with $V$ the unit cell volume and $N_e$ number of electrons in it) and the number of electrons in a given spin configuration (either $N_{e, \uparrow}$ or $N_{e, \downarrow}$). 
	These quantities are easily translated into the Wigner-Seitz radius, $r_s$ given by $\frac{4}{3}\pi r_s^3= \frac{V}{N_e}$ and the spin polarisation  $\zeta=\frac{|N_{e, \uparrow}-N_{e, \downarrow}|}{N_e}$.
	The discretisation of the Bloch wave-vectors introduced in \autoref{sec:plwaves}, however, requires also a specific choice for the $\mathbf{k}$-point sampling and for the symmetry of the simulation cell.

		Calculations were performed on a $\Gamma-$point  centred, uniform $\mathbf{k}$-point mesh of dimensions $N_k \times  N_k \times N_k $. 
	Furthermore, we applied simple cubic unit cells in all our simulations. 
	As already emphasized before results are independent of the choice of the unit cell, once $\mathbf{k}$-point convergence is reached. 
	We confirmed for instance that a face-centred cubic cell and a simple cubic cell result in the same RPA energy.
	Simple cubic cells are, however, often used for quantum Monte-Carlo simulations. 
	For instance, a $3\times 3\times 3$ $\mathbf{k}$-point grid with 2 electrons in the unit cell corresponds exactly to a  $3\times 3\times 3$ super cell with 54 electrons which was often used in QMC simulations, because the resultant electronic configuration has only fully occupied or entirely unoccupied one-electron orbitals \onlinecite{Lin2001}.

		The VASP code \cite{KresseVASP1, KresseVASP2} has been used for all calculations, which required a three-stage computational procedure. 
	The one-body Green's function variation was disregarded along the coupling constant integral, as discussed above.
	The solidity of this approximation will be checked by comparing the resulting correlation energy against the QMC benchmark. 
	For a given $(r_s,\zeta)$ the procedure to evaluate the correlation energy starts with a self-consistent calculation at the Kohn-Sham level.
	The plane-wave basis set is specified by the energy cutoff $E_{\rm cut}$ and includes plane waves with kinetic energy smaller than the given cutoff value. 
	This step is followed by a self-consistent evaluation of the quasi-particle energies at the $QP-GW_0$ level \cite{KresseGW, KresseGWa}.
	The screening potential $W_0$ is evaluated at the RPA level with orbitals and eigenvalues from the previous DFT calculation. 
	To ensure consistency of the results with previous calculations \cite{KresseSOSEX} for the evaluation of the correlation energy we set the cutoff for the response function's basis set (\textsc{encutgw}) and densities in the two electron integrals to the same value $E_{\rm cut}$.
	The frequency sampling has been carried out on a linear grid, specified by its maximum value, set equal to 1.5 $E_{\rm cut}$ for all $(r_s,\zeta)$. 
	The grid point density has been chosen as 0.1 for $r_s$=0.5, 0.8 and 0.2 for $r_s$=1.0 (collectively referred as the high density region), 
	then progressively increased to 0.6 and 1.0 in the intermediate ($r_s$=2.0, 3.0) and low densities ($r_s$=4.0, 5.0), respectively. 
	The evaluation of quasi-particle energies, necessary to specify the dressed propagator, has been performed iteratively.
	We set the number of iterations to three; the resulting quasi-particle energies have a mean accuracy of $\approx$ 1 mRy.
	These are the only quantities being updated (plane waves being exact eigenfunctions for the HEG).
	Thus, the one-body Green's function retains a simple one electron form.

		The BSE was recast as in \autoref{eq:RPAsq} and has been implemented in VASP; all virtual unoccupied orbitals spanned by the plane-wave basis were included. 
	The diagonalisation of \autoref{eq:RPAsq} is required for each value of the coupling constant, 
	whereas the Coulomb kernel (constant along the AC-path) is computed and stored once and for all at the beginning of the BSE step. 
	The integral in \autoref{eq:Ecfinal} is evaluated numerically with a Gauss-Legendre quadrature scheme.
	As few as 2 points are sufficient to produce results converged within the order of meV.

	
	\begin {figure*} [t]
	\centering
	\includegraphics [width=15.0cm] {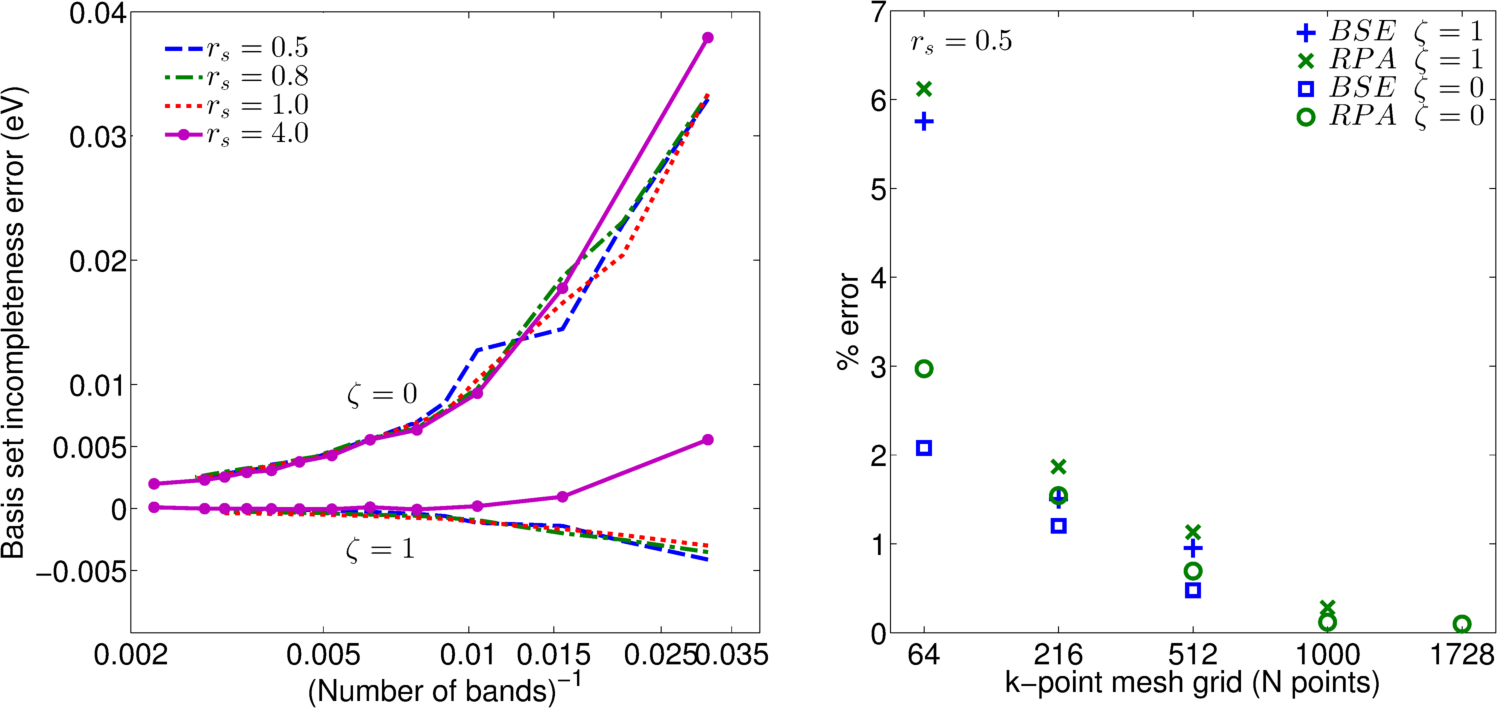}
	\caption{
	Left panel: convergence for the BSE correlation energy as a function of the total number of plane waves; the energy zero has been set equal to the linearly extrapolated complete basis energy limit for each value of $(r_s,\zeta)$. 
	Right panel: total energy  convergence as a function of the number of points included in the reciprocal space sampling. 
	Here the shown value is calculated by taking the difference between the total energy for any given $\mathbf{k}$-mesh and the value at the preceding sampling.}
	\label{Fig:conv}
	\end{figure*}

	\section{Results and Discussion}
	\label{sec:Results}
	
	\begin{figure}
	\includegraphics [width=7.5cm] {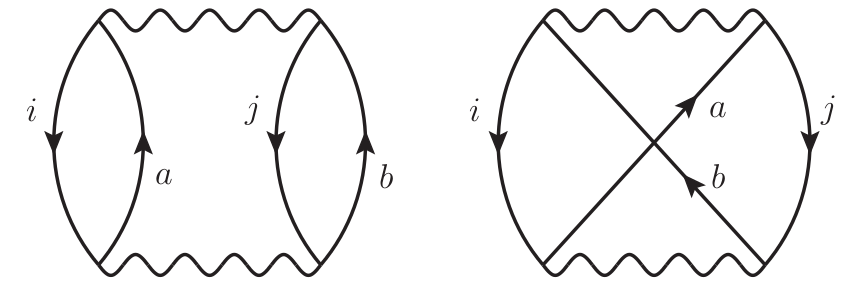}
	\caption{Direct (left) and exchange (right) diagrams contributing in second order to the correlation energy.}
	\label{Fig:secOrder}
	\end{figure} 
 
		We start by considering the correlation energy's convergence properties for a given $(r_s, \zeta)$ against the total number of plane waves. 
	The impact of the basis set incompleteness is shown in \autoref{Fig:conv} (left panel): for the set of densities considered, a linear convergence behaviour with the inverse of the number of bands is observed \cite{Harl2008b}.
	The complete basis set limit has been estimated for both the paramagnetic ($\zeta=0$) and ferromagnetic ($\zeta=1$) case with a linear regression. 
	For the ferromagnetic system a particularly fast convergence is observed as a function of the number of plane waves.
	To the best of our knowledge, an analytical wave-vector analysis is not available for arbitrary spin polarisations. 
	However, in Ref. \onlinecite{Gori-Giorgi2002} the pair correlation function limit for large wave-vectors is said to be determined by the kinks in the many electron wave function.
	For fully spin polarized systems these are completely absent, this then rationalises the fast convergence observed. 
	We will comment on this point in more detail below.

		The assessment of the total energy convergence with respect to the $\mathbf{k}$-point grid density is reported in the right panel of \autoref{Fig:conv}. 
	Here we report the total energy for any given $\mathbf{k}$-point mesh minus the value at the preceding $\mathbf{k}$-point sampling divided by the energy value obtained with the most accurate $\mathbf{k}$-point grid for a given level of theory.  
	For each $\mathbf{k}$-point grid, the total energy has been computed on the $GW_0$ reference state and includes the one-electron kinetic energy, the  exact exchange energy, a correction term related to the occurrence of fractional occupancies (Ref. \onlinecite{Harl2010a}), as well as the correlation contribution. 
	We include in the figure also the RPA energy convergence that provides an upper bound on the relative error. 
	The corresponding error on the total energy is less than 1\% for the ferromagnetic system and less than 0.5\% for the paramagnetic ground state.
	On the other hand, convergence of the correlation energy alone proves more difficult. 
	To estimate the correlation energy convergence with respect to both the basis set and the $\mathbf{k}$-point sampling we employ an extrapolation scheme adapted from Ref. \onlinecite{Klimes2014b}:
	\begin{align} 
		E^{\infty}_c(N_k \rightarrow \infty) \approxeq E_c^r(N_k \rightarrow \infty) - E_c^r(N_k) + E^{\infty}_c(N_k), 
	\label{eq:Ecextra}
	\end{align}
	where the superscript of the correlation energy $E_c$ indicates the dependence on the basis set.
	The symbol \textit{r} indicates that the calculations were performed using a small number of bands (32 bands here), whereas $\infty$ indicates that the results have been extrapolated to the infinite basis set limit.
	In practice, we found that it is sufficient to determine the basis set correction ($- E_c^r(N_k) + E^{\infty}_c(N_k)$) using $ N_k=3$, i.e. $3 \times 3 \times 3$ $\mathbf{k}$-points.

		To validate this somewhat involved approach, we first calculated the correlation energy starting from the Kohn-Sham ground state, i.e. RPA @ LDA.
	This should reproduce the previously published analytic results of Ref. \cite{PhysRev.176.589} for the paramagnetic system and of Ref. \cite{PhysRevB.23.5048} for the spin-polarised case. 
	We tested the agreement for the case $(r_s=1, \zeta=0)$ where we found it necessary to include up to $N_k=18$ points to reproduce the analytical results within 4 meV.
	Fewer $\mathbf{k}$-points ($N_k=16$) are necessary for the spin-polarised case to reproduce the data with a similar accuracy.
       We stress that the RPA results reported below have been calculated on the $GW_0$ reference and hence can not match the historical RPA correlation energies, owing to the renormalisation of the propagator lines in the RPA response function.

		Given the current computational limitations, we adapt the extrapolation scheme above to the case of post-RPA calculations with the replacement:  
	\begin{align} 
		E_c^r(N_k \rightarrow \infty) \approxeq E_c^r(N_k=8) + \Delta_{N_k}, 
	\label{eq:Ecextra2}
	\end{align}
	where the correlation energy is computed at the considered level of theory whereas the $\Delta_{N_k}$ correction corrects for the $\mathbf{k}$-point incompleteness error. This value is difficult to calculate accurately for the BSE. In previous work \cite{Marsman2009}, it was found that the second order exchange reduces the direct correlation energies by about 1/3. The situation is similar here, with exchange contributions reducing the correlation energy by roughly 30\%.
	In line with this, the $\mathbf{k}$-point errors are generally 1/3 smaller for BSE than for RPA (compare Fig. 2, but note that the left panel presents relative errors).
	Overall, it therefore seems sensible to determine the $\mathbf{k}$-point error using the computationally efficient RPA, but to reduce the RPA $\mathbf{k}$-point correction by  a factor  $\frac{2}{3}$ if exchange is included. 
	\begin {figure*} [t]
	\centering
	\includegraphics [width=15.0cm] {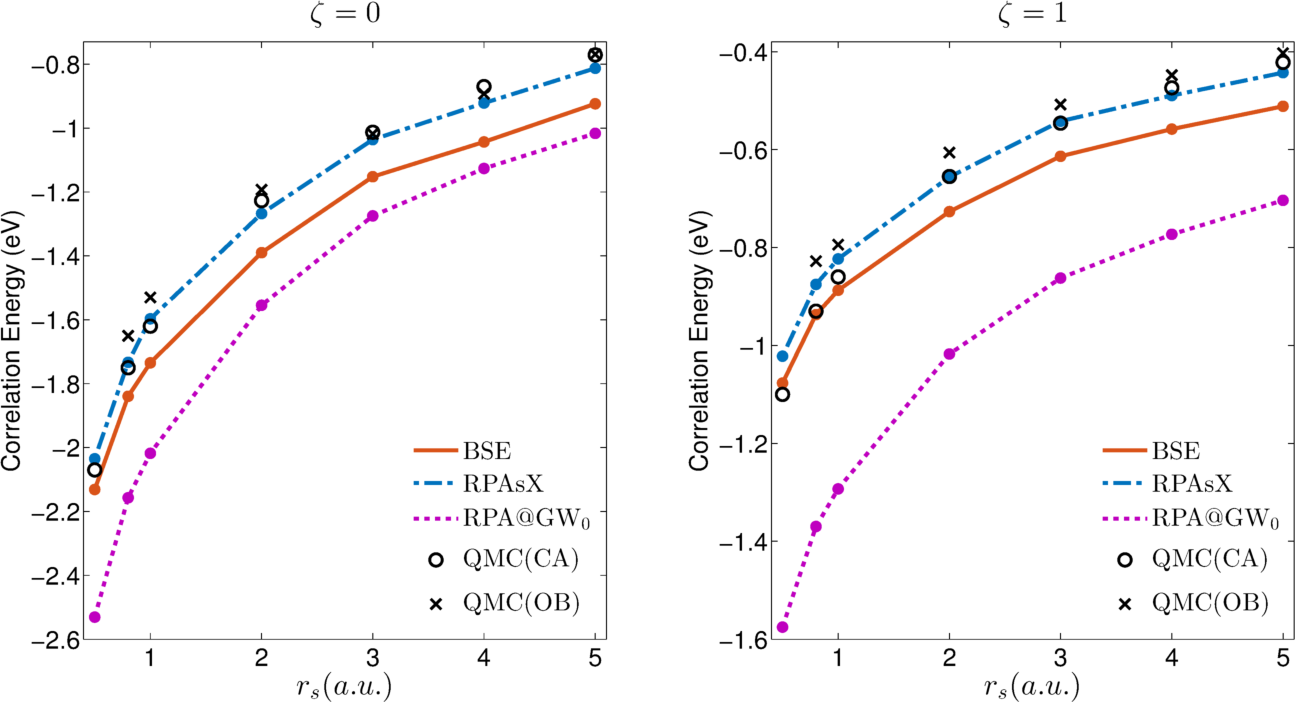}
	\caption{Values of correlation energy for the full BSE (solid line) and for the RPAsX approximation (dash-dotted line) as a function of the Wigner-Seitz radius for the paramagnetic (left) and ferromagnetic (right) HEG. 
	Results are compared against the QMC simulations in Ref. \onlinecite{Ceperley1980b} (round symbols) and in Ref. \onlinecite{Ortiz1994a} (crosses). 
	The random phase approximation on the same reference state (broken line) is also included.}
	\label{Fig:corrE}
	\end{figure*}
		
		The correlation energies computed for a set of $(r_s,\zeta)$ are shown in \autoref{Fig:corrE} by a solid line and compared against the Quantum Monte Carlo estimates by Ceperly and Adler (CA in the following) \cite{Ceperley1980b} (indicated by circles in the figure) and by Ortiz and Ballone (OB in the following) \cite{Ortiz1994a} (`$\times$' symbols in the figure). 
	Finite size effects impact also these benchmarks \cite{Fraser1996} and different extrapolation schemes, which do not necessarily fulfil the variational principle, have been employed. 
	Since we did not want to judge which calculation is more accurate, we compare to both results.
	The RPA@GW$_0$ estimate of the correlation energy (dotted line in \autoref{Fig:corrE}), is obtained as usually by including the bubble diagrams only, i.e.  excluding the ladder diagrams (A2) and (B2) in \autoref{Fig:BSEkernel}.
	The newly proposed RPA with screened exchange (dash-dotted line), where only bubbles and diagrams (B2) are included is also shown (see below).
	Starting our analysis from the full BSE calculations (solid line in \autoref{Fig:corrE}), in the paramagnetic case the CA results are reproduced with a mismatch of 0.06 eV for the highest density.
	The error progressively increases in the high density region and becomes significant with a deviation of $\approx$20\% for $r_s \ge$4.0 a.u..
	This pattern is also observed for the ferromagnetic system, with the absolute deviation, however, decreasing to $\approx$10\% at low densities.

		The general trend of the present BSE results can be rationalised as follows: 
	in the low density region the BSE makes a sizeable error, which is of unknown origin, but could be related to  
	the lack of particle-particle (or hole-hole) ladder diagrams to describe short range interactions \cite{Yasuhara1972}.
	Indeed, for the paramagnetic system (where short range interactions are most important) we observe that the BSE and RPA results are close in energy and both deviate from the QMC estimate. 
	In the high density region the BSE reproduces well the QMC estimates for both values of spin polarisation.  

		Short range interactions are small in the ferromagnetic case owing to the Pauli exclusion principle and are incorrectly included by the RPA. 
	This self-correlation error degrades the agreement between the QMC estimates and the RPA \cite{Paier2012a} (see left panel in \autoref{Fig:corrE}).
	The self-correlation error is cancelled in leading order by the BSE, as we will discuss now.
	In the low density region the large wave-vector contributions to the correlation energy become predominant \cite{Giuliani2005}. 
	At large wave-vectors, the screened interaction $W_0$ equals the bare interaction $V$, since screening is weak.
	The leading contribution to the BSE becomes then equal to the second order contribution in M{\o}ller-Plesset perturbation theory (MP2) given in Ref. \onlinecite{Jansen2010b}.
 	This can be shown by determining the leading term in the correlation part of the two body density matrix.
 	For the Hartree term and the RPA the equivalence has also been derived in Ref. \onlinecite{Marsman2009}.
	In second order, the correlation energy is then simply given by the standard textbook equation:
	\begin{align} 
		E^{(2)}_c=\frac{1}{2} \sum_{ijab} \frac{\langle ij|ab \rangle}{\Delta E_{ia} +\Delta E_{jb}} \big\{ \langle ab|ij \rangle - \langle ab|ji \rangle\big\}.
	\label{eq:Ec2ndO}
	\end{align}
        From this, two important observations follow.

	\noindent (i) The Hartree contribution $\langle ab | ij \rangle$ and the exchange term $ - \langle ab | ji \rangle$ cancel each other for $j=i$ or for $a=b$ for any spin orientation.
	This self-correlation error cancellation is obviously observed for all values of $\zeta$ since only same-spin electrons are affected by the self-correlation error. 
	In other words, MP2 is self-correlation free and this property is shared in second order by the BSE.
	This property is also observed by the SOSEX approximation \cite{KresseSOSEX}. 
	However, for SOSEX the corresponding second order diagram is introduced somewhat \textit{ad hoc} by anti-symmetrizing the direct-RPA coupled cluster amplitudes. 
	For the BSE, the diagrams are naturally included at the level of the two particle propagator and two body density matrix. 
	The BSE  should therefore improve upon the SOSEX, in particular, if static correlation effects are important\cite{Henderson2010}. 

        \noindent (ii) The expression for the correlation energy in second order also allows us to understand why the correlation energy converges so quickly with respect to the number of orbitals  for the ferromagnetic case.
        For a ferromagnetic system, only one spin component is present. 
	For large wave-vectors, the dominant terms to the correlation energy stem from $i=j$, and the direct term is exactly cancelled by the exchange term.
        This explains the fast convergence of the correlation energy with respect to the plane wave cutoff for the ferromagnetic case.
	For a paramagnetic system, where both spin components are present, it is possible to carry out the summation over the spin degrees of freedom implicitly present in the spin-orbital representation above. 
	This leads to a factor $2^l$, with $l$ being the number of fermionic loops in the diagrams in \autoref{Fig:secOrder}. 
        Hence, if \autoref{eq:Ec2ndO} is evaluated for spatial orbitals \cite{Paldus1975}, a factor four enters for the direct term, whereas a factor two is present for the exchange term.
        Thus the two terms do not cancel each other in the long wave-length limit, more precisely the direct term is reduced by a factor two by the exchange term. 
	This causes the slow convergence for the non-spin polarized system. 
	
		Given the relevance of the HEG as a model for condensed matter, it is not surprising that other research groups have recently evaluated its properties as well.
	For instance, the use of the renormalised ALDA kernel with the inclusion of correlation effects seems to reproduce exactly the HEG correlation energy in the limit of very low densities ($r_s>$ 10) \cite{Patrick2015}.
	However larger deviations are observed in the "metallic" density region ($r_s \in [0.5,5.0]$).
	Also, a recent study by the de Gironcoli's group \cite{Colonna2014a} has addressed the role of correlations in the HEG by employing exact exchange (EXX) TDDFT. 
	A feature of this method is to include first order changes of the one-particle propagator due to exact exchange, as well as particle-hole exchange interactions in the two-point response function (due to adiabatically switching on the exact exchange between electrons). 
	This response function is then used to form the exchange kernel as proposed by G\"{o}rling \cite{Gorling1998}. 
	An issue that this approach faces is the occurrence of imaginary frequencies when the response function is diagonalised (see Eq. (10) in \cite{Colonna2014a}). 
	Although it is possible to circumvent this problem by including the ladder diagrams (A2) and (B2) only to first order in the response function, this also spoils the results to some extent \cite{Colonna2014a}.
	Other difficulties in the EXX-TDDFT method stem from the inversion of the non-interacting response function to evaluate the exchange kernel (see Eq. (7) in \cite{Bleiziffer2012}) and on the critical dependence on the basis set size \cite{Hirata2001}. 

		Also in the present BSE approach instabilities at lower electron densities can be present (as in our case for $r_s \ge$5). These are witnessed as imaginary frequencies $\Omega$ appearing in \autoref{eq:RPAsq}. 
	The instabilities can cause the matrices $\mathbf{A+B}$ or $\mathbf{A-B}$ to have negative eigenvalues. 
	If both matrices have negative eigenvalues, it has been argued that the ground state is unstable with respect to particle-hole excitations \cite{Jørgensen1981}. 
	In the present case, however, we find that only the $\mathbf{A+B}$ matrix has negative eigenvalues, for both the $GW_0$ and LDA reference states. 
	This finding is consistent with a recent theoretical investigation \cite{Uimonen2015} which has confirmed the presence of negative-frequency modes in the BSE polarisation propagator.

		In our case the instabilities originate from the particle-hole ladder diagrams, as also witnessed in the case of the already mentioned TDDFT calculations. 
	To resolve this problem, we suggest to {\em disregard} all particle-hole ladder diagrams in the $\mathbf A$ matrix--- specifically those shown in (A2) of \autoref{Fig:BSEkernel} 
	---when solving the BSE equation. 
	In the following, we  will refer to this approximation as RPA with screened exchange (RPAsX).  

		We believe this choice to be sensible for the following reasons. 
	First, the $\mathbf{A}$ matrix does not contribute to the correlation energy in the perturbation expansion to second order \cite{Angyan2011a, Heßelmann2011b}.  
	Thus removing the diagrams (A2) will only change the correlation energy in third order; in second order, for instance, we still recover the exact MP2 energy (if the screened exchange kernel is replaced by the bare one).
	The approximation is also closely related to the AC-SOSEX method \cite{KresseSOSEX} but improves upon it. 
	In the AC-SOSEX the conventional RPA polarisation propagator is evaluated. 
	As opposed to the direct RPA, the polarisation propagator is then contracted against the $\mathbf{B}$ matrix (containing the sum of the direct Coulomb interaction and the bare exchange interaction, diagrams (B1) and (B2) in \autoref{eq:Ecfinal}) \cite{Jansen2010b}.
	In RPAsX, the propagator now includes, as it should, the exchange term in the $\mathbf{B}$ matrix (B1 as well as B2). 
 	Furthermore, the exchange terms (B2) now include a screened exchange interaction instead of a bare interaction, which effectively mimics higher order ladder diagrams. 
	It is clear from \autoref{Fig:corrE} that this approximation is particularly successful for the HEG, as it completely prevents the occurrence of unstable solutions and yields excellent agreement with the QMC data for both spin polarisation values
 	as shown in \autoref{Fig:corrE} by a dash-dotted line. 
	We observe a slight upward shift of the correlation energy compared to the full BSE approach, which leads to undercorrelation in comparison with CA by 0.03, 0.02 eV for $r_s$=0.5 and 0.8, 1.0 a.u. respectively.
	For $r_s \ge$2 the RPAsX estimate of the correlation energy remains within 6\% of the CA estimate for $\zeta=0$ (it overcorrelates to a greater extent in comparison with OB). 
	For the spin-polarised case RPAsX lies within the range of values spanned by the different QMC simulations in the high density region.
	As the density is lowered there is a progressive increase of the correlation energy in RPAsX, which still compares very favourably with CA values, reaching a maximum mismatch of 20 meV for the lowest density considered. 
	RPAsX correlation energy is mostly parallel to the  BSE results, but moves closer to the exact results.  
	We admit that the very good agreement with the QMC results must be fortuitous to some extend, since the RPAsX still neglects particle-particle and hole-hole ladder diagrams.
	But obviously the neglect of these diagrams cancels the also neglected  particle-hole ladder diagram at least for the HEG.

		Finally, we consider a partially spin polarised system. 
	The non-interacting response function is diagonal in the spin basis and this property is conserved by the BSE kernel, because opposite spin components interact only through the direct diagrams (A1) and (B1) in \autoref{Fig:BSEkernel}. 
	As it was mentioned, increasing the spin polarisation decreases the number of electrons able to interact at short range.
	To assess the change introduced in the correlation energy $E_c$ by modifying the fraction of short-ranged \textit{vs} long-range interactions, the spin enhancement function $\Upsilon(\zeta; r_s)$ is commonly introduced:
	\begin{align}
	\label{eq:Y}
	\Upsilon(\zeta ; r_s)=\frac{E_c(\zeta, r_s)-E_c(0, r_s)}{E_c(1, r_s)-E_c(0, r_s)}.
	\end{align} 
	
	\begin {figure} [h]
	\centering
	\includegraphics [width=7.3cm] {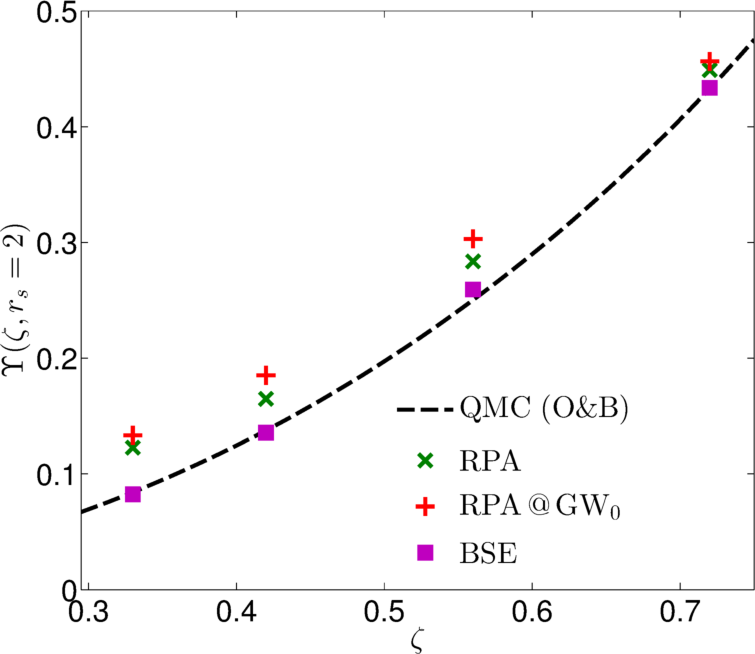}
	\caption{The spin enhancement function $\Upsilon$ as a function of the fractional polarisation at density $r_s$=2. Calculations were performed on a 6$\times$6$\times$6 $\mathbf{k}$-point grid with complete basis set extrapolation.}
	\label{Fig:PartPolar}
	\end{figure}

	\noindent In \autoref{Fig:PartPolar} we report the spin enhancement function computed at $r_s$=2 versus the spin polarisation $\zeta$ for various levels of theory.
	Results are compared against the Perdew-Wang interpolation formula \cite{PhysRevB.45.13244} with parameters estimated by Ortiz and Ballone \cite{Ortiz1994a} (dashed line). 
	RPA results are evaluated on the DFT reference state (represented in the figure by the `$\times$' symbols): there is a noticeable departure  from the QMC estimate, with the RPA overestimating the short range correlations, as expected.  
	Changing reference state  (RPA on $GW_0$ indicated by `+' symbols in the figure) actually increases the mismatch with the QMC estimate. 
	The BSE results, shown by squares in the figure, reproduce exceedingly well the benchmark calculations for all values of $\zeta$ considered.
	Given the high computational cost required by these calculations we do not report RPAsX for fractional spin polarisations. 
	However, since the BSE makes a more sizeable error than RPAsX (see \autoref{Fig:corrE}) and yet it reproduces $\Upsilon$ very accurately, it is reasonable to assume that RPAsX will also be very accurate for fractional values of $\zeta$.
	
	
	\section{Conclusions}
		In this study we have evaluated the correlation energy for the homogeneous electron gas by calculating the fluctuation contributions from the Bethe-Salpeter equation.
	The implementation of the BSE has taken advantage of the time inversion symmetry, converting it into an Hermitian quadratic eigenvalue problem.
 	Both the resonant and the antiresonant contributions to the particle-hole interaction have been included in the BSE (going beyond the so-called Tamm-Dancoff approximation). 
	
		The BSE kernel has been set up consistently with Hedin's $GW_0$ approximation.
	This level of theory has been employed also to describe the system's electronic structure, taken as a starting point for the ensuing BSE calculation. 
	The occurrence of unstable solutions using the full Bethe-Salpeter kernel has prompted us to seek an approximation scheme, still able to compare favourably with QMC references. 
	The approximation  proposed, called RPAsX, is \textit{by construction} consistent with MP2 and reproduces the QMC correlation energy very well for values of $r_s \ge$ 1.0 \textit{a.u.}.
	We certainly plan to test this approximation for a wider class of systems and materials. 
	There are however several obstacles that need to be solved before this approach can be applied routinely.
	The most important one is that for the HEG changes related to changes of the one-particle Green's function and one particle orbitals can be neglected (compared \autoref{eq:dEdl}).
	This is not the case for real materials, where self-consistency of the orbitals \cite{PhysRevLett.96.226402} or changes of the one particle Green's function when going to the interacting case are very relevant \cite{Klimes2015}. In order to apply the RPAsX to real systems it will be necessary to relax the approximations made here.
	With this provision, we are confident that this approach can be extended to realistic solid state and molecular systems.

	\section*{Acknowledgments}
	Financial support from the Austrian Science Found (FWF) for the SFB ViCoM project no. F41 is kindly acknowledged.
	
	\appendix
	\section{Derivation of Eq. 5}
	\label{appx:G2}

	\begin {figure*} [t!]
		\centering
		\includegraphics [width=15.9cm] {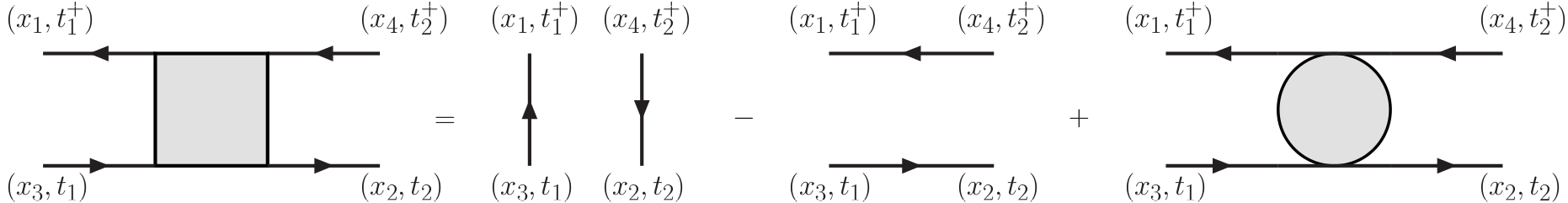}	
		\caption{Graphical representation for the two-body Green's function. We represent the time ordering $t_1=t_2$ which corresponds to having two electron-hole pairs interacting instantaneously}
		\label{Fig:G2}
	\end{figure*}

	In this appendix we derive the expression for the correlation energy in terms of the polarisation propagator.	
      The correlation energy involves the two-body density matrix and, as shown in the main text in \autoref{eq:dEdl}, is given by
        \begin{widetext}
	\begin{align}
		\left( \frac{dE_\lambda}{d \lambda} \right )_{\rm{2B}} \equiv \dot{E}_\lambda^c =  \frac{1}{2} \int dt_1\int \prod_{i=1}^4 d\mathbf{x}_i  
		\delta (\mathbf{x}_3 - \mathbf{x}_1) \delta(\mathbf{x}_4 - \mathbf{x}_2) v(\mathbf{x}_1,\mathbf{x}_2) 
		 \langle \psi^\dagger (\mathbf{x}_4, t^+_1) \psi^\dagger (\mathbf{x}_3, t_1) \psi (\mathbf{x}_1, t^+_1) \psi (\mathbf{x}_2, t_1) \rangle_\lambda \nonumber. \\
		\label{eq:dEdl2B}
	\end{align}
        \end{widetext}
	The four field operators in this expression suggest that one can make a connection to Green's function theory, specifically to the two particle Green's function.
	We also note that the expression above only involves equal time limits (all involved time points are equal).
	To make a connection to Green's function theory, we first need to introduce a time dependence in the field operators by defining the time evolution operator in the interaction picture (as it is commonly done in many body perturbation theory):
	$\hat{\psi}(1)=e^{i\mathcal{H}_0 t_1} \hat{\psi}(\mathbf{x}_1)e^{-i\mathcal{H}_0 t_1}$
	and  
	$\hat{\psi}^\dagger(2)=e^{i\mathcal{H}_0 t_2} \hat{\psi}^\dagger(\mathbf{x}_2)e^{-i\mathcal{H}_0 t_2}$.
	The $n$-body Green's operator is as usually defined as: 
	\begin{align}
		\label{eq:Gn}
		\mathcal{G}_{n} & (1,2,...,n;  1',2',..., n')= \nonumber \\ 
		& \frac{1}{i^n} \mathcal{T} 
		\{\hat{\psi}(1) \hat{\psi}(2) \dots \hat{\psi}(n)
		\hat{\psi}^\dagger(n') \dots \hat{\psi}^\dagger(2') \hat{\psi}^\dagger(1')\}  
	\end{align}
	where $\mathcal{T}$ is the time ordering operator. 
	It is worth pointing out that the operators $\mathcal{G}_n$ do not depend on the value of the coupling constant $\lambda$. 
	For $n=1$ the corresponding (greater) Green's function is obtained by evaluating the expectation value of the Green's operator for the ground-state wave function at coupling $\lambda$, $\Psi_\lambda$. 
	Since we are only interested in equal time limits here, we can restrict the second time to fulfil $t'=t+0^+=t^+$:
	\begin{align}
		\langle \mathcal{G}_{1} (\mathbf{x}_1,t^+; \mathbf{x}_2, t) \rangle_\lambda &\equiv \nonumber \\ 
		G_{1,\lambda} (\mathbf{x}_1,t^+; \mathbf{x}_2, t)&= i\langle \psi^\dagger (\mathbf{x}_2,t) \psi(\mathbf{x}_1,t^+) \rangle_\lambda \label{eq:G1} 
	\end{align} 
	At equal time limits, the Green's function can be also related to the density matrix
	$\langle \hat{n} \rangle_\lambda \equiv n_\lambda(\mathbf{x}, \mathbf{y}) =- i G_{1,\lambda} (\mathbf{x}, t ; \mathbf{y}, t)$.

	For the two particle Green's function, it is standard texbook material \cite{Gross1991} (commutator relations or Wick's theorem) to show that it satisfies:
	\begin{align}
		G_{2,\lambda}&(\mathbf{x}_1, t^+_1, \mathbf{x}_2,t_2;  \mathbf{x}_3,t_1, \mathbf{x}_4,t^+_2) \nonumber \\ 
		= & -\langle \psi^\dagger (\mathbf{x}_3, t_1)  \psi (\mathbf{x}_1, t^+_1)\rangle_\lambda
		\langle \psi^\dagger (\mathbf{x}_4, t^+_2)  \psi (\mathbf{x}_2, t_2)\rangle_\lambda \nonumber \\
		& +
		\langle \psi^\dagger (\mathbf{x}_4, t^+_2)  \psi (\mathbf{x}_1, t^+_1)\rangle_\lambda
		\langle \psi^\dagger (\mathbf{x}_3, t_1)  \psi (\mathbf{x}_2, t_2)\rangle_\lambda \nonumber \\
		& + \label{eq:G2dec}
		\langle \psi^\dagger (\mathbf{x}_4, t^+_2) \psi^\dagger (\mathbf{x}_3, t_1) \psi (\mathbf{x}_1, t^+_1)\psi (\mathbf{x}_2, t_2)\rangle_{\lambda},
	\end{align}
	where the last term on the r.h.s. is the correlation part of the two-body Green's function that is also present in the AC fluctuation-dissipation theorem (ACFDT) correlation energy \autoref{eq:dEdl2B}.
	The labelling convention for four-point quantities is given in \autoref{Fig:G2} and it is consistent with Ref. \onlinecite{Starke2012} up to a time inversion.
	Since the correlation energy involves only equal time limits, we again set $t_1=t_2=t$. 
	This allows us to rewrite \autoref{eq:G2dec} in terms of the ordinary density operators $\hat{n}$ and their expectation values: 
	\begin{align}
		G_{2,\lambda}&(\mathbf{x}_1, t^+, \mathbf{x}_2,t;  \mathbf{x}_3,t, \mathbf{x}_4,t^+) \nonumber \\ 
		= & -\langle \psi^\dagger (\mathbf{x}_3, t)  \psi (\mathbf{x}_1, t^+)\rangle_\lambda
		\langle \psi^\dagger (\mathbf{x}_4, t^+)  \psi (\mathbf{x}_2, t)\rangle_\lambda \nonumber \\
		& + n_\lambda(\mathbf{x}_4,\mathbf{x}_1)n_\lambda(\mathbf{x}_3,\mathbf{x}_2)  
		- \langle \hat{n}(\mathbf{x}_4,\mathbf{x}_1) \hat{n}(\mathbf{x}_3,\mathbf{x}_2) \rangle_\lambda \nonumber \\
		= & -\langle \psi^\dagger (\mathbf{x}_3, t)  \psi (\mathbf{x}_1, t^+)\rangle_\lambda
		\langle \psi^\dagger (\mathbf{x}_4, t^+)  \psi (\mathbf{x}_2, t)\rangle_\lambda \nonumber \\
		& - \langle \delta \hat{n}(\mathbf{x}_4,\mathbf{x}_1) \delta \hat{n}(\mathbf{x}_3,\mathbf{x}_2) \rangle_\lambda,
		\label{eq:G2dn}
	\end{align} 
	where in going to the last line, we have introduced the density fluctuation operator $ \langle\delta \hat{n}\rangle_\lambda \equiv \langle \hat{n} \rangle_\lambda -  n_\lambda$.
	The last term in \autoref{eq:G2dn} can be identified with the polarisation propagator (also called density-fluctuation density-fluctuation 
 response function), \textit{i.e.} $P_\lambda\equiv\langle\delta\hat{n}\delta\hat{n}\rangle_\lambda$.
	One can then rearrange \autoref{eq:G2dn} into the usual relation connecting the two particle propagator and the polarisation propagator \cite{Stefanucci2013}:
	\begin{align}
		P_\lambda&(\mathbf{x}_1, t^+, \mathbf{x}_2,t;  \mathbf{x}_3,t, \mathbf{x}_4,t^+) = \nonumber \\
		& - G_{2,\lambda}(\mathbf{x}_1, t^+, \mathbf{x}_2,t;  \mathbf{x}_3,t, \mathbf{x}_4,t^+) \nonumber \\ 
		& + G_{1,\lambda}(\mathbf{x}_1, t^+ ; \mathbf{x}_3,t)G_{1,\lambda}(\mathbf{x}_2, t ; \mathbf{x}_4,t^+).
		\label{eq:G2Pl}
	\end{align}

	Using \autoref{eq:G2dec} and \autoref{eq:G2Pl}, one can rewrite the correlation part of $ G_{2,\lambda}$ in the ACFDT equation using the polarisation propagator and obtain the following compact equation for the correlation energy:
	\begin{align}
		\dot{E}^{\rm{c}}_\lambda =
		 \frac{1}{2} \int \prod_{i=1}^4 d\mathbf{x}_i  \, V & \left [G_{1,\lambda}(\mathbf{x}_1, t^+; \mathbf{x}_4, t^+)  
		G_{1,\lambda}(\mathbf{x}_2, t; \mathbf{x}_3, t) \right. \nonumber \\
		& \left. - P_\lambda ( \mathbf{x}_1, t^+, \mathbf{x}_2, t; \mathbf{x}_3, t, \mathbf{x}_4, t^+)  \right]  
		\label{eq:E_xc}
	\end{align}
	From \autoref{eq:E_xc} it is easy to separate out the exchange contribution:
	\begin{equation} \label{eq:EX_lambda}
	\begin{split}
		\dot{E}^{\rm{x}}_\lambda &= 
		 - \frac{1}{2} \int \prod_{i=1}^4 d\mathbf{x}_i  \delta(\mathbf{x}_3-\mathbf{x}_1) \delta(\mathbf{x}_4-\mathbf{x}_2) \nonumber \\
		& \qquad \qquad v(\mathbf{x}_1,\mathbf{x}_2) n_\lambda(\mathbf{x}_2,\mathbf{x}_3) n_\lambda(\mathbf{x}_1,\mathbf{x}_4) \nonumber \\
		&= - \frac{1}{2} \int d\mathbf{x}_1 d\mathbf{x}_2   
			 v(\mathbf{x}_1,\mathbf{x}_2)  n_\lambda(\mathbf{x}_2,\mathbf{x}_1) n_\lambda(\mathbf{x}_1,\mathbf{x}_2). 
	\end{split}
	\end{equation}
	The remaining contribution in \autoref{eq:E_xc} can be replaced into \autoref{eq:Ec} to form the correlation energy assuming that the one body contributions (contained in the exchange part) remain constant along the AC path.
	\autoref{eq:EcP} is then obtained by Fourier transforming with respect to the infinitesimal time difference.

	\section{Spectral representation for the polarisation propagator} 
	\label{appx:deriv}

	The starting point to derive the spectral representation for the interacting polarisation propagator $P_\lambda(\omega)$ is to invert \autoref{eq:Pi} in the particle-hole basis:
	\begin{align}
		\mathbf{P}_\lambda(\omega) = \left [ \mathbf{P}_0^{-1}(\omega) -\mathbf{I}_\lambda \right ]^{-1}.
		\label{eq:invBSE}
	\end{align}
	We note that for real matrices an analogous but formally somewhat different presentation can be found in Furche \cite{Furche2001a}.
	Poles in the polarisation propagator will be located at those frequencies that fulfill the condition $\operatorname{det} \{ \mathbf{P}_0^{-1}(\omega) -\mathbf{I}_\lambda \} =0 $.
	Given the spectral representation for $\mathbf{P_0(\omega)}$ in \autoref{eq:L0SR} it is straighforward to construct its inverse, both being diagonal matrices:
	\begin{align}
	\mathbf{P}_0^{-1}(\omega) &= 
		\begin{pmatrix}
			-\Delta E_{ia} & \mathbf{0} \\
			\mathbf{0} & \Delta E_{ia}
		\end{pmatrix}
		+ \omega
		\begin{pmatrix}
			\mathbf{1} & \mathbf{0} \\
			\mathbf{0} & \mathbf{-1}
		\end{pmatrix} \nonumber \\
	&= \mathbf{D} + \omega \mathbf{\Delta}.
	\end{align} 
	In going to the last line we have introduced the supermatrices $\mathbf{D}$ and $\mathbf{\Delta}$ for convenience.
	The polarisation propagator can then be compactly expressed as: 
	$	\mathbf{P}_\lambda(\omega) = \left [ \mathbf{D} - \mathbf{I}_\lambda + \omega \mathbf{\Delta} \right ]^{-1} 
		= - \left[ \mathbf{M}_\lambda - \omega \mathbf{\Delta}\right ]^{-1}$,  
	where the square supermatrix $\mathbf{M}_\lambda$ corresponds to the one introduced in \autoref{eq:RPA}.
	Since the determinant is invariant under similarity transformations, it is convenient to introduce the matrix:
	\begin{align}
		\mathbf{Z}_\lambda=
		\begin{pmatrix}
		\mathbf{X}_\lambda & \mathbf{Y}_\lambda^* \\
		\mathbf{Y}_\lambda & \mathbf{X}_\lambda^*
		\end{pmatrix}
	\end{align} 
	that fulfils the generalised eigenvalue equation  
	\begin{align}
	\mathbf{M}_\lambda \mathbf{Z}_\lambda = \mathbf{\Delta}\mathbf{Z}_\lambda \mathbf{\Omega}_\lambda
	\label{eq:EVPapp}
	\end{align}
	which is nothing but a re-statement of \autoref{eq:RPA} taking into account the complex conjugate solution.
	The matrix $\mathbf{Z}_\lambda $ satisfies the "symplectic" normalisation condition $\mathbf{Z}_\lambda \mathbf{\Delta} \mathbf{Z}_\lambda^\dagger=\mathbf{\Delta}$;
	this constraint, together with $\mathbf{\Delta}$ being idempotent, are sufficient to show that $\mathbf{Z}_\lambda$ is unitary, and that the condition 
	\begin{align}
	\label{eq:DZD}
	\mathbf{\Delta} \mathbf{Z}_\lambda^\dagger \mathbf{\Delta} = \mathbf{Z}_\lambda^{-1} 
	\end{align}
	holds.
	
		The relation in \autoref{eq:DZD} can be exploited to construct the spectral representation of the polarisation propagator starting from \autoref{eq:EVPapp}:
	\begin{align}
	\{\mathbf{M}_\lambda - \omega \mathbf{\Delta} \} \mathbf{Z}_\lambda & = \mathbf{\Delta} \mathbf{Z}_\lambda (\mathbf{\Omega}_\lambda - \omega) \\
	\Rightarrow \mathbf{M}_\lambda - \omega \mathbf{\Delta} & = \mathbf{\Delta} \mathbf{Z}_\lambda (\mathbf{\Omega}_\lambda - \omega)\mathbf{\Delta} \mathbf{Z}_\lambda^\dagger \mathbf{\Delta}. 
	\end{align}
	Inverting the right hand side of the previous equation one finally obtains:
	\begin{align}
	\label{eq:LSR}
	\mathbf{P}_\lambda (\omega) = \mathbf{Z}_\lambda \left ( \omega - \mathbf{\Omega}_\lambda \right )^{-1} \mathbf{\Delta}\mathbf{Z}_\lambda^\dagger. 
	\end{align} 
	The presence of a simple pole analytic structure for both polarisation propagators entering \autoref{eq:EcP} is of fundamental importance because it allows the straightforward frequency integration necessary to evaluate the correlation part of the two body density matrix $\mathcal{P}_\lambda$.


\bibliography{BSE_Jellium_sub}
\end{document}